\documentclass[aip,pop,reprint]{revtex4-1}

\draft 

\usepackage{amsmath}
\usepackage{xfrac} 

\newcommand{\espic}{ESPIC}

\begin{document}


\title{Kinetic Electron and Ion Instability of the Lunar Wake
  Simulated at Physical Mass Ratio}



\author{Christian Bernt Haakonsen}
\email[]{chaako@mit.edu}
\author{Ian H. Hutchinson}
\email[]{ihutch@mit.edu}
\author{Chuteng Zhou}
\email[]{ctzhou@mit.edu}
\affiliation{Plasma Science and Fusion Center, \
  Massachusetts Institute of Technology}


\date{\today}

\begin{abstract}
  The solar wind wake behind the moon is studied with 1D electrostatic
  particle-in-cell (PIC) simulations using a physical ion to electron
  mass ratio (unlike prior investigations); the simulations also apply
  more generally to supersonic flow of dense magnetized plasma past
  non-magnetic objects. A hybrid electrostatic Boltzmann electron
  treatment is first used to investigate the ion stability in the
  absence of kinetic electron effects, showing that the ions are
  two-stream unstable for downstream wake distances (in lunar radii)
  greater than about three times the solar wind Mach
  number. Simulations with PIC electrons are then used to show that
  kinetic electron effects can lead to disruption of the ion beams at
  least three times closer to the moon than in the hybrid
  simulations. This disruption occurs as the result of a novel wake
  phenomenon: the non-linear growth of electron holes spawned from a
  narrow dimple in the electron velocity distribution. Most of the
  holes arising from the dimple are small and quickly leave the wake,
  approximately following the unperturbed electron phase-space
  trajectories, but some holes originating near the center of the wake
  remain and grow large enough to trigger disruption of the ion
  beams. Non-linear kinetic-electron effects are therefore essential
  to a comprehensive understanding of the 1D electrostatic stability
  of such wakes, and possible observational signatures in ARTEMIS data
  from the lunar wake are discussed.
\end{abstract}

\pacs{52.30.-q, 52.35.Fp, 52.35.Mw, 52.35.Qz, 52.35.Qz, 52.35.Sb,
  52.35.Tc, 52.65.Rr, 95.30.Qd, 96.20.-n, 96.25.Qr, 96.25.St,
  96.50.Ci} 

\maketitle 

\section{Introduction}
\label{sec:intro}

The supersonic flow of the solar wind past the moon and other objects
without global magnetospheres is a topic of continued theoretical and
experimental interest. The ongoing ARTEMIS mission \cite{Russell2014}
is studying the solar wind in the vicinity of the moon, with targeted
3D hybrid simulations being used \cite{Wiehle2011} to interpret the
observed fluctuations \cite{Halekas2011}. More generally, simulations
of dense magnetized plasma flow past non-magnetic objects has wide
applicability, ranging from probes in laboratory plasmas to asteroids
and the moons of Saturn \cite{Khurana2008,Simon2009}.

Solar wind flow past the moon leads to an elaborate lunar plasma
environment and wake \cite{Halekas2011a}. At distances beyond a few
lunar radii the main dynamics of interest are those associated with
the ambipolar plasma expansion into the void trailing the moon, and in
particular any resulting instabilities. This process has been studied
with fully kinetic simulations in 1D
\cite{Farrell1998,Farrell2008,Birch2001b} and 2D \cite{Birch2002}, as
well as with hybrid simulations in 1D \cite{Israelevich2012}, 2D
\cite{Travnicek2005}, and 3D
\cite{Kallio2005,Wang2011,Wiehle2011,Holmstrom2012}. There are also 2D
kinetic simulations which ignore the magnetic field of the solar wind
\cite{Guio2005,Kimura2008}, but these are less relevant to the moon
given its large size relative to the gyroradius scales in the solar
wind.

Despite the wide range of prior simulations, none have so far
accurately captured the 1D electrostatic phenomena which can occur in
the wakes of large non-magnetic objects. Birch and Chapman
\cite{Birch2001b} attributed the fluctuations they (and previously
Farrell et al. \cite{Farrell1998}) observed to electron beam
instability. However, Hutchinson \cite{Hutchinson2012} showed that the
unphysical ion to electron mass ratio used in those simulations
artificially enhances the energy scale of electron--electron
instability, indicating that a realistic mass ratio is needed to
accurately model such instabilities in the wake. The present work
confirms the importance of using a realistic mass ratio, discovering a
novel wake phenomenon where electron holes grow from the small energy
scale of the electron--electron instabilities to a scale where they
can disrupt the counterstreaming ion beams in the wake. This
observation has already inspired theoretical investigations into the
underlying mechanism \cite{Hutchinson2015}, which show that the
changing density in the wake can act as a drive for hole growth.

Kinetic electron effects are thus shown to be important in the flow of
magnetized plasmas past objects much larger than the plasma Debye
length. The case of supersonic flow is treated here, but the failure
of Boltzmann or other hybrid electron treatments to capture the
relevant electrostatic phenomena may well extend to cases with slower
plasma flow: counterstreaming ions will also be present for slower
flow, and the electrons will still see a varying potential and local
density, so the main differences are geometrical and related to
boundary conditions at the collecting object. Whether these
differences are enough to render the kinetic electron effects
unimportant is a topic that should be investigated, as it may be
relevant to Mach probes and other means of invasive plasma
measurements.

This paper is organized as follows: The 1D electrostatic simulation
approach used is described in Section~\ref{sec:1dSimulations}.
Simulations using Boltzmann electrons are presented in
Section~\ref{sec:boltzmann}, showing the onset of ion two-stream
instability; the unstable (but mass-ratio dependent) electron
distributions associated with the Boltzmann-simulated wake are also
illustrated. Fully kinetic simulations are presented in
Section~\ref{sec:kinetic}, where some of the electron holes formed
early in the wake are seen to grow and lead to disruption of the ion
beams much closer to the moon than in the Boltzmann-electron
simulations. Possible observational signatures of electron holes and
ion-beam disruption in ARTEMIS data are discussed in
Section~\ref{sec:observations}, and some concluding remarks given in
Section~\ref{sec:conclusions}.

\section{1D Electrostatic Simulation Method}
\label{sec:1dSimulations}

The lunar wake simulation is reduced to 1D through the same approach
taken in prior such studies \cite{Farrell1998,Farrell2008,Birch2001b}:
by following the plasma in the frame moving with the solar wind, where
the perpendicular motion is restricted by the magnetic field. What is
simulated is then the parallel motion of electrons and ions along an
unperturbed magnetic field line as a function of time. For a static
wake this time-dependence can be translated to a spatial dependence by
following the trajectory of the field line, and a single 1D simulation
thus applies to a wide range of solar wind flow velocities and
magnetic field orientations with respect to the moon
\cite{Hutchinson2012} (in suitably scaled coordinates).

For ease of comparison with the prior studies
\cite{Farrell1998,Farrell2008,Birch2001b} the relative motion of the
solar wind and the moon is taken to be perpendicular to the magnetic
field and have speed $v_\mathrm{sw}=25c_\mathrm{s}$. Here
$c_\mathrm{s}=\sqrt{\sfrac{T_\mathrm{e}}{m_\mathrm{i}}}$ is the (cold
ion) sound speed, which is equal to the ion thermal velocity since
$T_\mathrm{e}=T_\mathrm{i}$. The distance behind the moon is thus
$x=v_\mathrm{sw}t$, and the position parallel to the magnetic field is
denoted $y$. The particular choice of solar wind speed does not enter
into the simulations, only the representation of the results in the
figures, so other speeds can be accommodated by a simple scaling of
$x$. Note that the convention for $x$ used herein differs from that
used in References~\onlinecite{Hutchinson2012,Hutchinson2015} by a
factor of the (cold ion) Mach number
$\sfrac{v_\mathrm{sw}}{c_\mathrm{s}}=25$.

A 1D electrostatic particle-in-cell (PIC) code, referred to here as
\espic, has been developed to carry out the simulations. The ions and
electrons are evolved with a standard leap-frog scheme at time steps
small enough to resolve the electron motion, and their charge
distributed to the two adjacent nodes of the mesh using the
cloud-in-cell approach \cite{Birdsall1991}. At each time-step the
finite-difference Poisson equation is solved for the potential using a
direct tridiagonal method, yielding a constant electric field between
any two mesh nodes to accelerate the particles. The potential at the
boundaries is set to zero with homogeneous Dirichlet boundary
conditions, and particles are injected there according to the assumed
Maxwellian background distributions.

\espic\ can be run in parallel to accommodate the large number of
particles and time-steps needed to resolve the wide ranges of relevant
length and time scales. Even so, large scale-separations such as in
the case of the lunar wake present a challenge: the lunar radius
$R_\mathrm{M}$ is $1730\,\mathrm{km}$, while the ion thermal
gyroradius is around $40\,\mathrm{km}$ and the electron and ion Debye
lengths are around $7\,\mathrm{m}$ \cite{Ogilvie1996,Halekas2005}. To
resolve the Debye length in a domain many lunar radii across would
thus require $\sim\,$$10^7$ grid-points, allowing for a few
grid-points per Debye length to reduce discretization error. That is
not currently feasible using a realistic ion to electron mass ratio,
but extrapolation based on simulations at larger (but still small)
Debye lengths can yield insight.

For such 1D electrostatic simulations to be applicable, the main
influence of the magnetic field on the particles must be to restrict
their motion in the directions perpendicular to it. Further, the
collecting object (e.g., the moon) must have small apparent extent in
the direction of flow, such that the ions do not move much as the
field line traverses the object. This can either be because the flow
is sufficiently supersonic or because the object has a large aspect
ratio, and in either case the detailed shape of the object doesn't
matter; its main influence is to leave a region depleted of plasma
behind it. The parallel dynamics of the ambipolar plasma expansion
into that depleted region then dominate the solution, and can be
captured in a 1D domain by suddenly removing the ions and electrons
from the part of that domain crossed by the object; this is the
approach taken in \espic.

Despite not directly influencing the ions, the shape (and surface
charge) of the object could still affect the (much faster)
electrons. For example, in the case of the lunar wake the electron
thermal speed is faster than the solar wind flow (for a realistic mass
ratio), so it is not very accurate to model the passage of the moon by
a sudden removal of electrons. However, the simultaneous removal of
both species provides a convenient way to remove an equal number of
electrons and ions, which avoids unphysical charge build-up associated
with inaccurate object boundary conditions (especially important for
simulations with short Debye length). Further, as will be shown in
Section~\ref{sec:boltzmann}, the initial electron disturbance (and
thus any inaccuracy associated with it) quickly propagates out of the
domain in simulations with a realistic mass ratio. Thus, the way of
implementing the (2D) passage of the moon in the 1D \espic\
simulations is thought to accurately capture the relevant effects on
the part of the wake of interest in the present work.

The main differences between the \espic\ simulations and prior 1D
kinetic simulations \cite{Farrell1998,Farrell2008,Birch2001b} are
finer phase-space resolution (up to $2\times10^4$ grid points and
$10^9$ particles), the use of a physical mass ratio, and the absence
of periodic boundary conditions. Preventing the wake from interacting
with itself through the periodic boundaries set a lower limit on the
domain size in the prior simulations, and for a physical mass ratio
that constraint would be unmanageable. Care must still be taken to
avoid reflection of potential structures from the boundaries, but that
is much less restrictive in terms of domain size than the periodic
boundaries; if need be absorbing potential boundary conditions could
be added allow even smaller domains.

\begin{figure*}[t!]
\centering
\includegraphics[width=0.7\linewidth, bb=0 0 325.82 191.13]{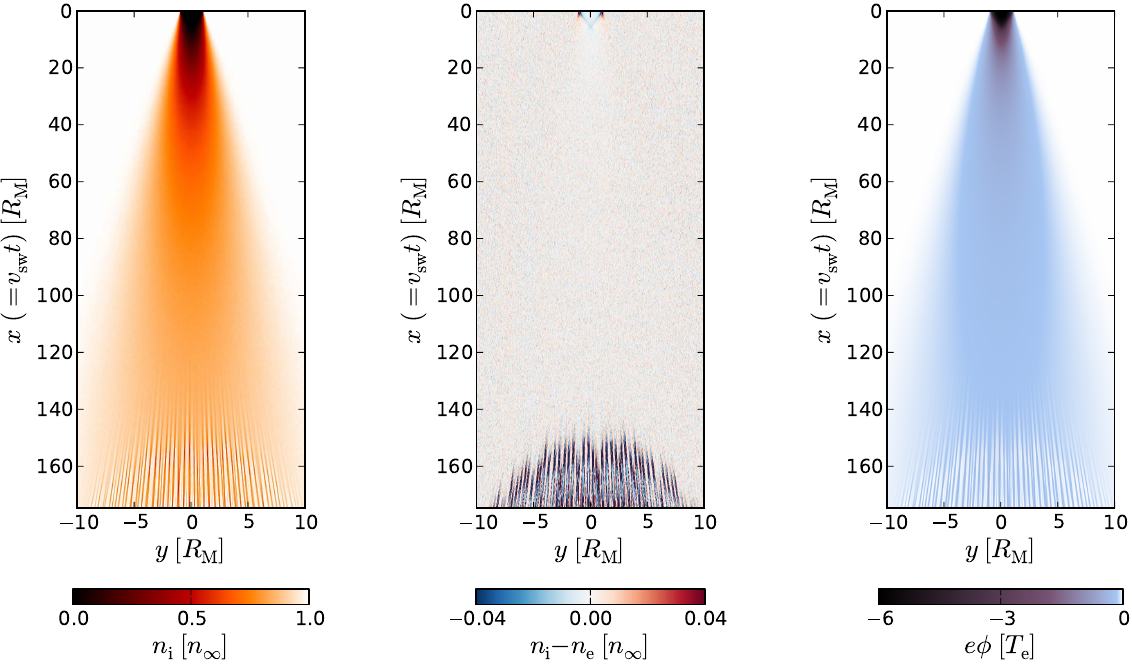}
\caption{Ion density $n_\mathrm{i}$ (in units of solar wind density
  $n_\mathrm{\infty}$), normalized charge density $n_\mathrm{i}\!-\!
n_\mathrm{e}$, and potential $\phi$ of the wake simulated with
Boltzmann electrons for $\lambda_\mathrm{De}=0.02R_\mathrm{M}$ (with
the illustrative plasma flow speed $v_\mathrm{sw}=25c_\mathrm{s}$).}
\label{fig:boltzmannWakeStructure}
\end{figure*}

For small Debye length $\lambda_\mathrm{D}$, a random initialization
of particles in the PIC method yields unphysically large fluctuations
in the initial potential. This is especially true at long
wave-lengths, as the amplitude of potential fluctuations of wavelength
$\lambda$ can be shown to scale with
$\left(\sfrac{\lambda}{\lambda_\mathrm{D}}\right)^{\sfrac{3}{2}}$ for
Gaussian counting statistics. For the same number of particles, the
amplitude of the potential fluctuations can be dramatically reduced by
using a {\em quiet start}\cite{Birdsall1991}, where particle positions
and velocities are initialized more uniformly than in a true random
start; a quiet start is used in \espic.

\section{Boltzmann Electron Simulations}
\label{sec:boltzmann}

Hybrid simulation approaches involve the use fluid or other
non-kinetic treatments of electrons to greatly reduce the
computational cost of simulations, which then only need to resolve the
ion time-scales as opposed to the (typically much shorter) electron
ones. One such approach is to assume that the electron density along a
magnetic field line satisfies a Boltzmann relation
\begin{equation}
n_\mathrm{e} =
n_\mathrm{e0}\exp\left(\frac{e\phi}{T_\mathrm{e}}\right)\, ,
\label{eq:boltzmann}
\end{equation}
which (for instance) it would exactly along a static, monotonically
decreasing (repulsive) potential $\phi$ from some point with a
stationary Maxwellian electron distribution of density
$n_\mathrm{e0}$. Such {\em Boltzmann electrons} are implemented in
\espic\ by solving the non-linear Poisson equation arising from using
Equation~\ref{eq:boltzmann} for the electron density, and are used in
the simulations throughout this section (even when moving electrons
kinetically to examine their distribution).

\begin{figure}
\centering
\includegraphics[width=\linewidth, bb=0 0 203.82 177.13]{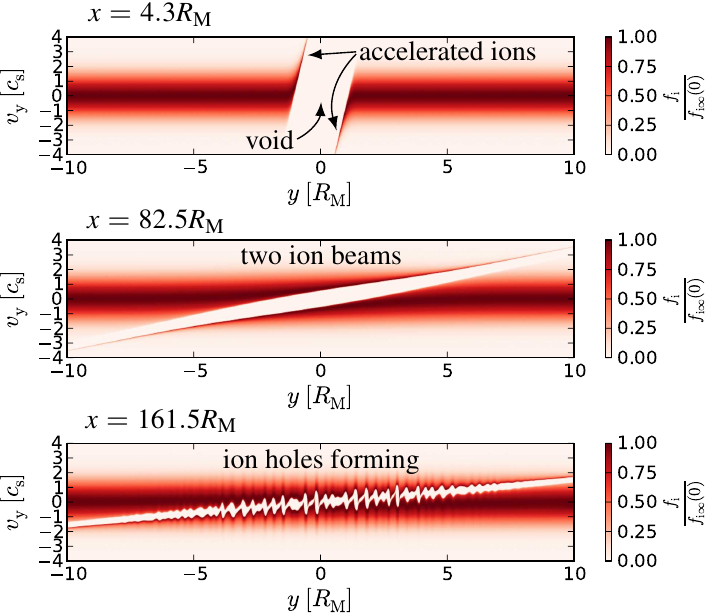}
\caption{Ion distribution at three different times (and thus $x$) in
  the simulation shown in Figure~\ref{fig:boltzmannWakeStructure}.}
\label{fig:boltzmannIonPhaseSpace}
\end{figure}

\begin{figure*}[t!]
\centering
\includegraphics[width=0.7\linewidth, bb=0 0 325.82 204.13]{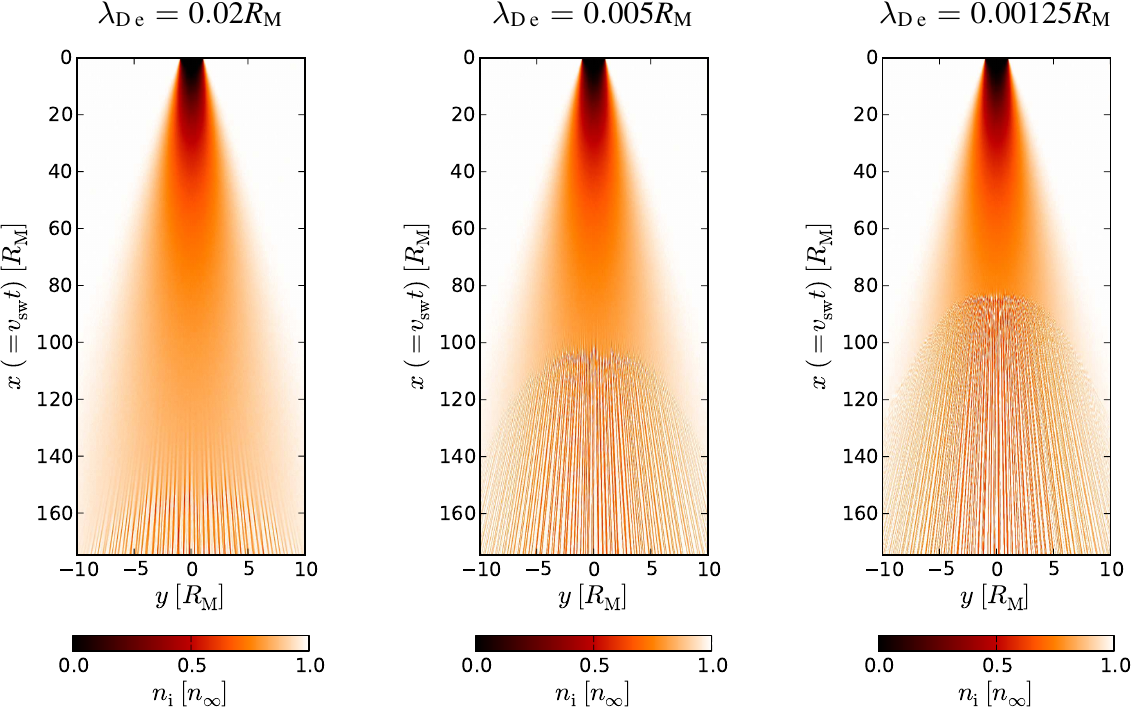}
\caption{Ion density in the wake for Boltzmann-electron simulations at
  three different $\lambda_\mathrm{De}$.}
\label{fig:boltzmannInstabOnset}
\end{figure*}

The wake structure resulting from a simulation with Boltzmann
electrons and a solar wind electron Debye length
$\lambda_\mathrm{De}=0.02R_\mathrm{M}$ is shown in
Figure~\ref{fig:boltzmannWakeStructure}. As discussed in
Section~\ref{sec:1dSimulations} a solar wind speed
$v_\mathrm{sw}=25c_\mathrm{s}$ is used to facilitate comparison with
prior 1D kinetic simulations
\cite{Farrell1998,Farrell2008,Birch2001b}. The Boltzmann-electron
simulation captures the main features of the initial plasma expansion
into the void behind the moon seen in the prior simulations: A
negative potential in the density-depleted region approximately
balances the electron and ion densities, with significant charge
density seen only in the transition from unperturbed plasma to the
strongly density-depleted region. Later in the wake, for $x\gtrsim 140
R_\mathrm{M}$, the simulation shows rapid ion (and charge) density
variations (vertically elongated striations in the plots) resulting
from ion two-stream instability (sometimes referred to as the ion--ion
[acoustic] instability \cite{Stringer1964,Gary1987}).

Figure~\ref{fig:boltzmannIonPhaseSpace} shows the ion phase space for
the Boltzmann-electron simulation at three different
times/positions. The main initial response of the ions is that they
are accelerated towards the void by the negative potential
structure. Later in the wake free-streaming of ions is the dominant
process, distorting the initial phase space void such that the
density-depletion is shallower but more spread out. This sets up a
configuration of two ion beams across most of the wake, and the
velocity gap between the two beams gradually shrinks. Eventually the
the ion two-stream instability grows to a non-linear stage, where
large ion holes form and give rise to the ion density variations seen
in Figure~\ref{fig:boltzmannWakeStructure}.

Simulations with shorter Debye length show instability-related density
variations earlier in the wake, as can be seen in
Figure~\ref{fig:boltzmannInstabOnset}. The large-scale structure of
the wake is not sensitive to the change in Debye length, so the
earlier appearance of the non-linear features of the two-stream
instability is caused mainly by the faster growth associated with the
higher plasma frequency at shorter Debye length; it takes a
significant time for unstable perturbations to grow large enough to be
visible when the Debye length is not infinitesimal. Extrapolating to
very small Debye-lengths indicates that the ion beams become
two-stream unstable when $t\approx
3\sfrac{R_\mathrm{M}}{c_\mathrm{s}}$, which corresponds to a
distance of $\sim\,$$75R_\mathrm{M}$ behind the moon for the
particular solar wind speed used in the figures.

\begin{figure}
\centering
\includegraphics[width=\linewidth, bb=0 0 404.00 188.00]{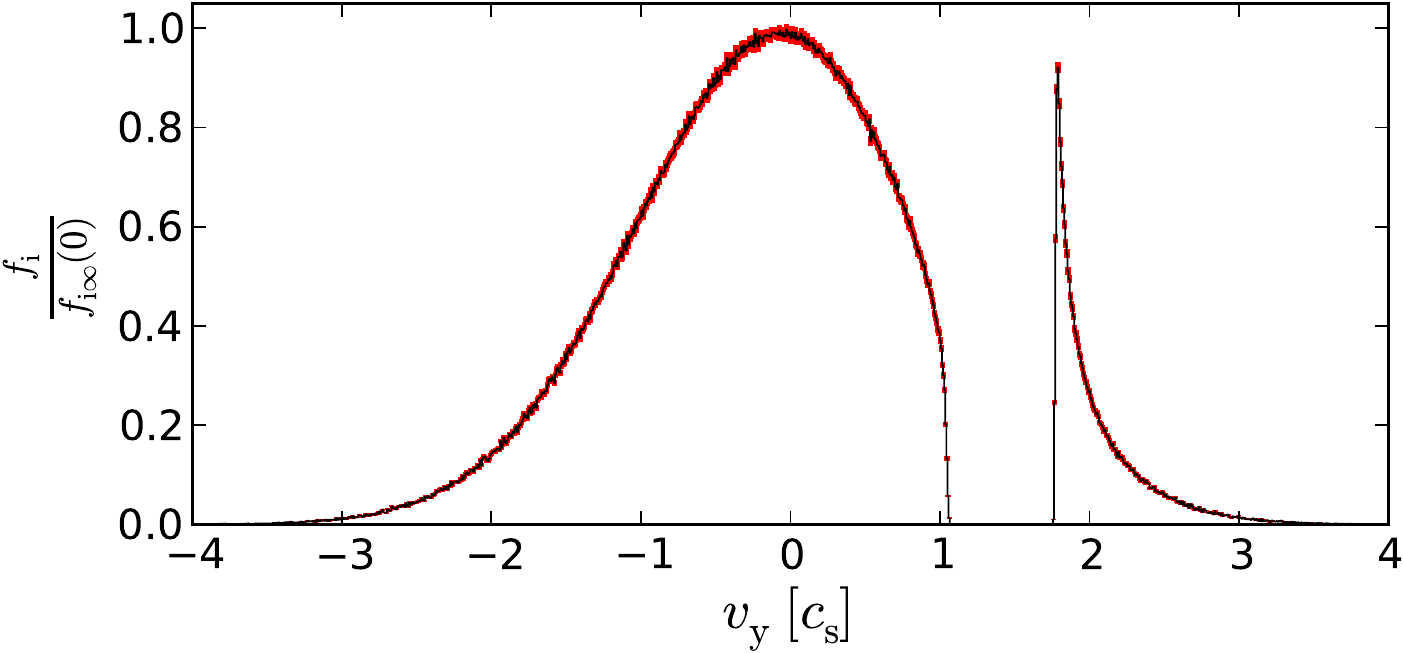}
\caption{Ion distribution at $x=82.5R_\mathrm{M}$ and
  $y=5R_\mathrm{M}$ in Figure~\ref{fig:boltzmannIonPhaseSpace}, with
  Gaussian counting errors for each bin shown in red.}
\label{fig:exampleDistribution}
\end{figure}

\begin{figure}
\centering
\includegraphics[width=0.45\linewidth, bb=0 0 201.00 404.00]{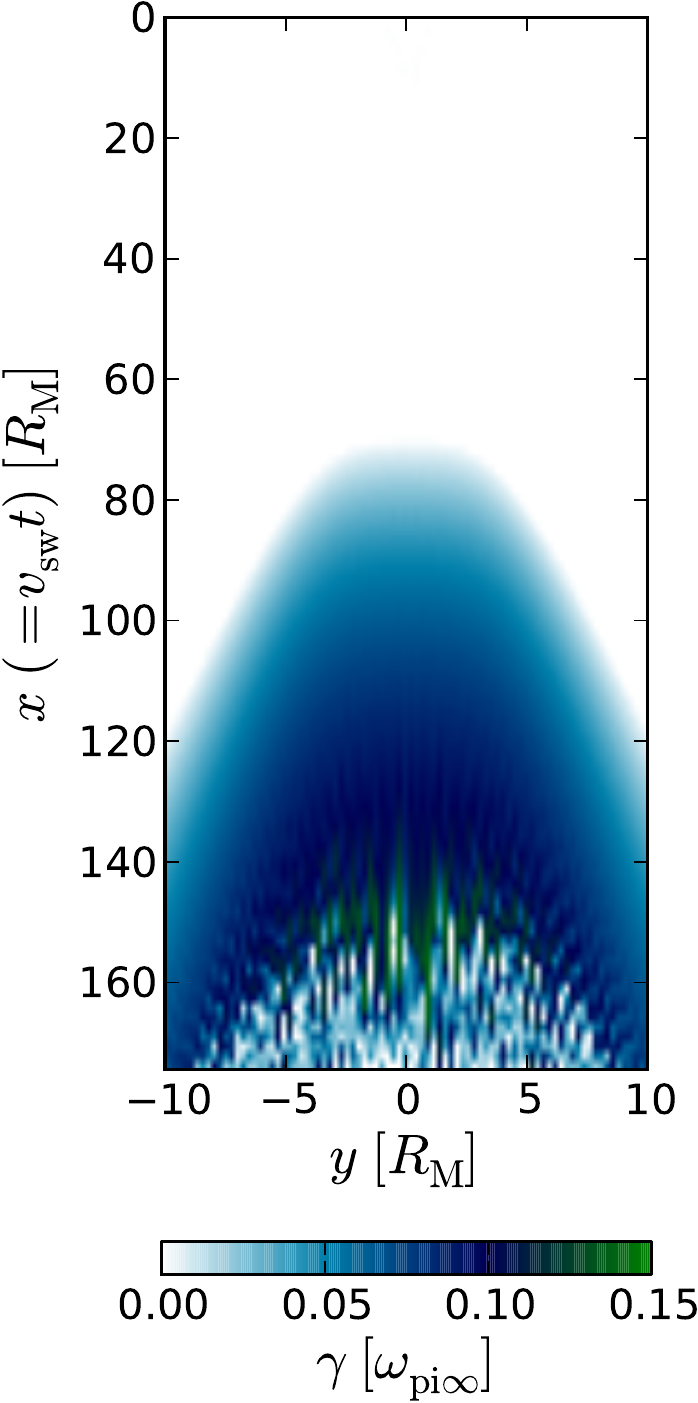}
\caption{Maximum linear ion-instability growth rate in wake for
  simulation shown in Figure~\ref{fig:boltzmannWakeStructure}, in
  units of the solar wind ion plasma frequency
  $\omega_\mathrm{pi\infty}$.}
\label{fig:linearStability}
\end{figure}

Traditionally, Maxwellian beams are considered when studying the
two-stream instability (e.g., Ref.~\onlinecite{Stringer1964}), and
cold ion beams are found to be unstable when their separation is less
than about twice the sound speed. However, as illustrated in
Figure~\ref{fig:exampleDistribution} the ion beams in the simulated
wake are not Maxwellian, and have sharp edges associated with the
initial void. To include enough particles to resolve the fine features
in velocity-space, large spatial bins of width $4\lambda_\mathrm{De}$
are used when storing the distributions for the following stability
analysis, leading to a small amount of smoothing.

The electrostatic ion stability at each location in the wake has been
analyzed by calculating numerically the complex Fourier-transformed
susceptibility of the (slightly further smoothed) ion distribution at
that location together with Maxwellian electrons, in a 1D linearized
Vlasov--Poisson system, and setting it equal to zero. The
ion--electron instability is suppressed in this calculation by
artificially setting the imaginary part of the electron susceptibility
to zero; this is appropriate because a Boltzmann electron treatment
excludes electron Landau damping or growth. Important deviations of
the electron distribution from a Maxwellian will be discussed later,
but are not considered here as attention is restricted to the ion
two-stream instability. The growth rate $\gamma$ of the most unstable
mode at each location is shown in Figure~\ref{fig:linearStability},
which clearly supports the interpretation of
Figure~\ref{fig:boltzmannInstabOnset} and the inferred instability
onset. The position chosen for the example distribution in
Figure~\ref{fig:exampleDistribution} can be seen to lie close to the
boundary between stable and unstable distributions.

If the imaginary part of the electron susceptibility is included, weak
ion--electron instability appears across most of the wake. This
instability is likely associated with the sharp edges of the ion
beams, and appears to be the instability discussed by Farrell et
al. \cite{Farrell1997} The Boltzmann-electron simulations of this
section do not capture that instability, but even in the fully kinetic
simulations in Section~\ref{sec:kinetic} it does not appear to play a
large role.

\begin{figure}
\centering
\includegraphics[width=\linewidth, bb=0 0 304.82 184.13]{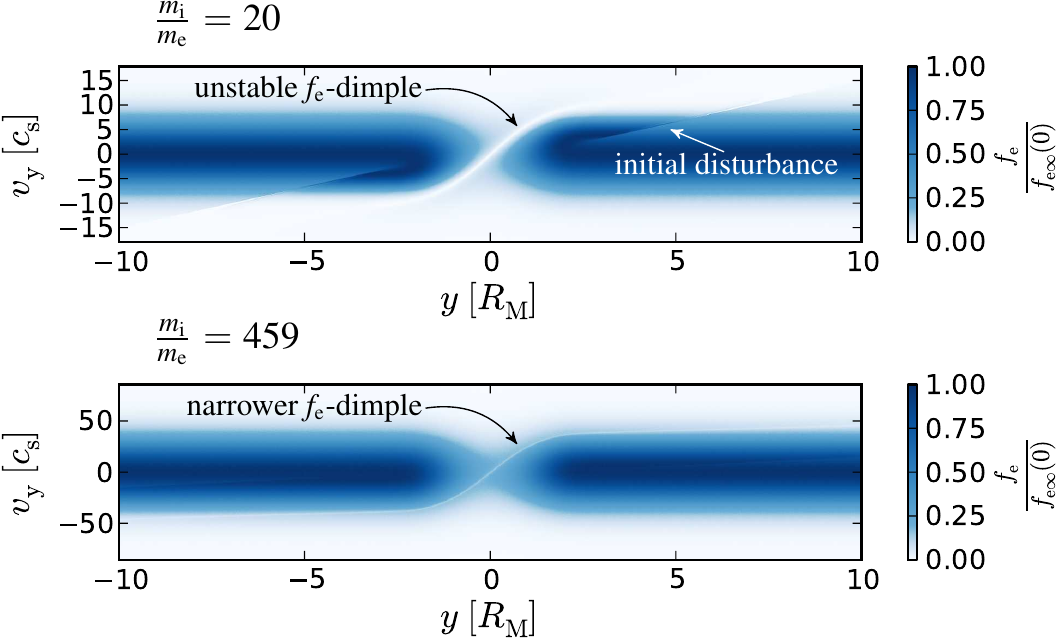}
\caption{Electron distribution from potential evolution in
  Boltzmann-electron simulation, at time corresponding to
  $x=15R_\mathrm{M}$, for two different mass ratios.}
\label{fig:boltzmannDimple}
\end{figure}

Before moving to fully kinetic simulations, it is useful to examine
the electron distributions that arise from the potentials of the
Boltzmann-electron simulations. Hutchinson \cite{Hutchinson2012}
showed that for a model potential evolution the electron distribution
develops an unstable depression (a {\em dimple}) at a
position-dependent velocity. Such a dimple is indeed seen in the
electron distributions from the Boltzmann-electron simulations,
examples of which are shown in Figure~\ref{fig:boltzmannDimple}. The
dimple is more pronounced for more artificial mass ratios, leading
Birch \& Chapman \cite{Birch2001b} to treat the electron distribution
as two electron beams in their stability analysis for
$\sfrac{m_\mathrm{i}}{m_\mathrm{e}}=20$. For a realistic mass ratio the
dimple only affects a very narrow velocity range of an otherwise
largely Maxwellian electron distribution, so treating the distribution
as two beams is not appropriate. Also seen in
Figure~\ref{fig:boltzmannDimple} is the initial disturbance to the
electron distribution from the (somewhat unphysical) sudden removal of
electrons from the computational domain. The disturbance propagates
out of the domain very quickly at realistic mass ratios, so any
inaccuracy associated with the 1D treatment of the passage of the moon
is unlikely to affect the parts of the wake of interest in the present
work.

\section{Fully Kinetic Simulations}
\label{sec:kinetic}

Protons are 1836 times more massive than electrons, providing a lower
bound for the true ion to electron mass ratio
$\sfrac{m_\mathrm{i}}{m_\mathrm{e}}$ in a typical plasma. Simulations
run at such mass ratios are computationally expensive, so a common
approach is to use an unphysically small mass ratio in an attempt to
capture the essential physics of the scale separation without paying
the full computational cost. This was done in prior PIC simulations of
the lunar wake \cite{Farrell1998,Farrell2008,Birch2001b,Birch2002},
which used the highly unphysical mass ratio
$\sfrac{m_\mathrm{i}}{m_\mathrm{e}}=20$. Later, Hutchinson
\cite{Hutchinson2012} showed that an unphysical mass ratio greatly
enhances the degree to which the lunar wake electron distribution is
unstable, and Hong et al. \cite{Hong2012} showed that an unphysical
mass ratio can significantly affect ion-driven beam instabilities in
ways that are not easily scaled to a physical mass
ratio. Acknowledging these findings, the present work uses simulations
at unphysical mass ratios only to illustrate phenomena observed in
simulations using the physical mass ratio for hydrogen.

\begin{figure}
\centering
\includegraphics[width=\linewidth, bb=0 0 208.82 191.13]{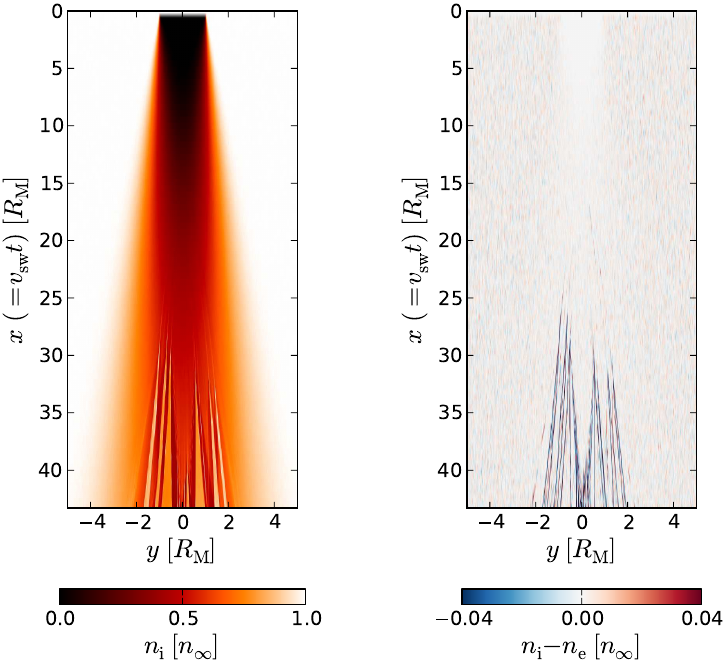}
\caption{Ion density $n_\mathrm{i}$ and normalized charge density
  $n_\mathrm{i}\!-\! n_\mathrm{e}$ of wake simulated with kinetic
  electrons for $\lambda_\mathrm{De}=0.0025R_\mathrm{M}$.}
\label{fig:kineticIonDensity}
\end{figure}

At the true mass ratio and $\lambda_\mathrm{De}=0.0025R_\mathrm{M}$,
treating both electrons and ions kinetically in \espic\ gives rise to
the ion density shown in Figure~\ref{fig:kineticIonDensity}.  Treating
the electrons kinetically gives rise to instability much earlier than
was seen in the simulations with Boltzmann electrons
(cf. Figure~\ref{fig:boltzmannInstabOnset}). Non-linear disruption of
the ion beams in this case occurs at
$t\approx1.1\sfrac{R_\mathrm{M}}{c_\mathrm{s}}$, which corresponds to
a distance of $\sim\,$$28R_\mathrm{M}$ behind the moon (at the
particular solar wind speed used for illustration). The nature of the
density variations is also different, with fewer, less periodic
density enhancements gradually widening.

\begin{figure}
\centering
\includegraphics[width=\linewidth, bb=0 0 204.82 176.13]{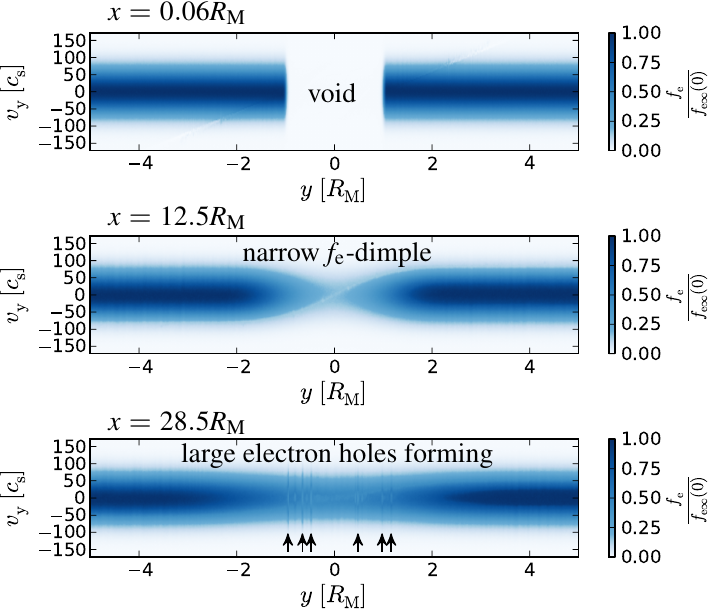}
\caption{Electron distribution at three different times (and thus $x$)
  in the simulation shown in Figure~\ref{fig:kineticIonDensity}.}
\label{fig:electronPhaseSpace}
\end{figure}

For the same simulation, the evolution of the electron distribution is
shown in Figure~\ref{fig:electronPhaseSpace}. Very early in the wake,
the only noticeable features are the initial void and the (faint)
outward propagating initial disturbance from the sudden removal of
electrons. Later in the wake a narrow dimple is seen, showing that the
dimple persists in some form also when self-consistently considering
electron instabilities
(cf. Figure~\ref{fig:boltzmannDimple}). Eventually, electron holes
with large velocity extent form and gradually widen in spatial extent,
corresponding to the density enhancements seen in
Figure~\ref{fig:kineticIonDensity}.

\begin{figure}
\centering
\includegraphics[width=\linewidth, bb=0 0 304.82 81.13]{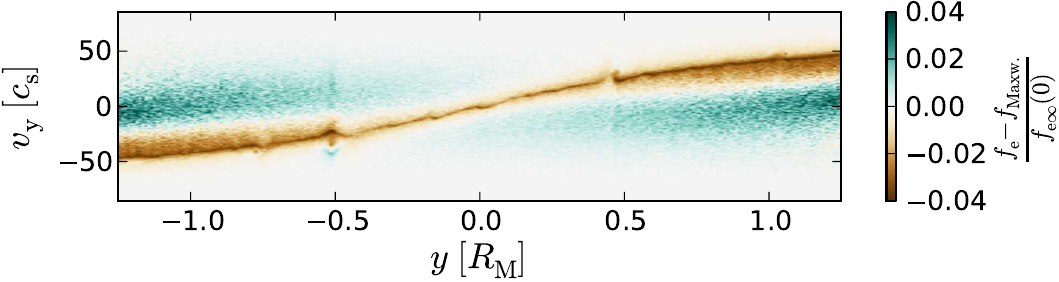}
\caption{Departure of electron distribution from a Maxwellian for a
  simulation with $\sfrac{m_\mathrm{i}}{m_\mathrm{e}}=459$ and
  $\lambda_\mathrm{De}=0.0025R_\mathrm{M}$.}
\label{fig:holeGrowthBeg}
\end{figure}

Because simulations at the true mass ratio have an extremely narrow
dimple and are somewhat noisy, a simulation using
$\sfrac{m_\mathrm{i}}{m_\mathrm{e}}=459$ (but the same Debye length)
is now used to illustrate the relevant electron phenomena. To better
visualize the dimple and holes, a Maxwellian of the same density (and
the solar wind electron temperature) is subtracted from the electron
distribution at each point in space. The result is shown in
Figure~\ref{fig:holeGrowthBeg}, for a time before electron holes with
large velocity-extent develop. Small hole-like perturbations are seen
along the dimple in the electron distribution, and examining
consecutive time-steps reveals that they convect outwards along the
dimple in phase space (without significantly perturbing the ions).

\begin{figure}
\centering
\includegraphics[width=\linewidth, bb=0 0 303.82 171.13]{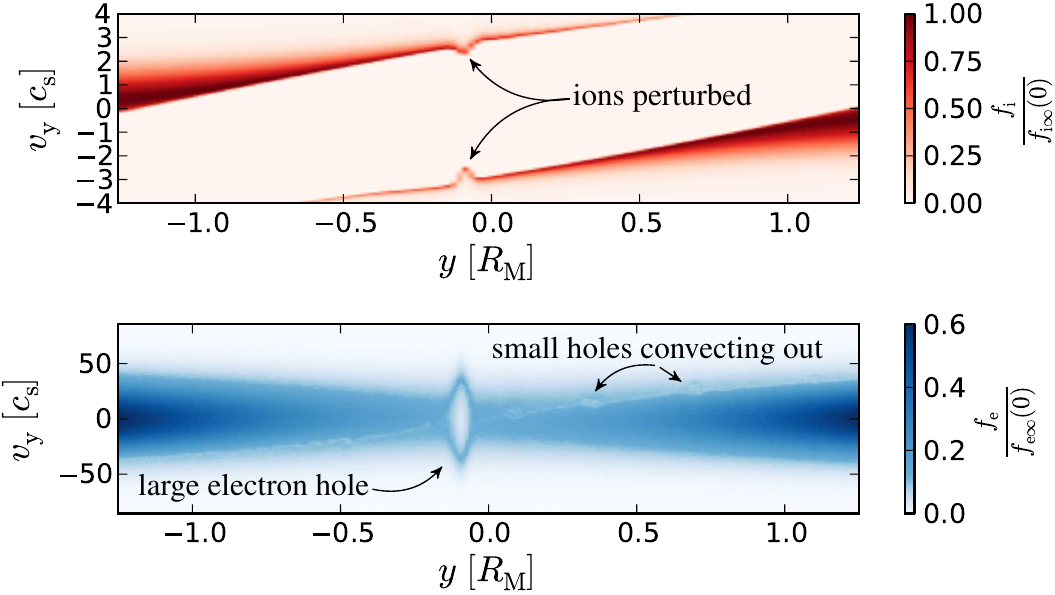}
\caption{Ion and electron distributions at a somewhat later time than
  in Figure~\ref{fig:holeGrowthBeg}.}
\label{fig:holeGrowthEnd}
\end{figure}

At a slightly later time, an almost stationary electron hole near
$y=0$ starts growing in velocity extent; the hole moves slowly
($v_\mathrm{y}\sim -c_\mathrm{s}$) towards negative $y$, not appearing
to accelerate to follow the dimple like the other holes. The ion and
electron distributions in the central region of the domain are shown
in Figure~\ref{fig:holeGrowthEnd}, for a time when the hole has grown
to a large enough extent to significantly perturb the ion beams. After
that time the ion beams are quickly disrupted (similarly to in prior
simulations \cite{Farrell1998,Birch2001b}), further widening the
electron hole in velocity as the gap between the ion beams closes. A
mechanism has been proposed to explain the growth of such holes
\cite{Hutchinson2015}, whereby the increasing local density as the
hole moves downstream drives an expansion of its velocity-extent.

Returning to the simulation at the true mass ratio, a careful
examination of the evolution of the electron distribution reveals that
the large electron holes seen in Figure~\ref{fig:electronPhaseSpace}
arise in the same way as the one discussed for the slightly reduced
mass ratio: small holes emerge from the dimple near $y=0$, move slowly
away from the dimple while growing in velocity-extent and remaining
centered at small $v_\mathrm{y}$, and then eventually expand to their
full velocity-extent as they disrupt the ion beams; the holes then
keep expanding in spatial extent as seen in
Figure~\ref{fig:kineticIonDensity}. The first seed of a large hole
emerges at a time corresponding to $x\approx14R_\mathrm{M}$, where the
local electron density is $0.21n_\infty$, and the holes grow until
they disrupt the ions around $x=28R_\mathrm{M}$, at which point the
local density is in the range $0.46n_\infty$ to $0.56n_\infty$
(depending on the hole position). For the same time-interval the total
velocity-separation of the ion beams shrinks from
$\sim\,$$5c_\mathrm{s}$ to $\sim\,$$3c_\mathrm{s}$. That density
increase and beam-separation decrease is sufficient for the proposed
hole growth mechanism \cite{Hutchinson2015} to lead to ion-beam
disruption, consistent with the simulations.

Because the electron holes do not remain exactly at $y=0$ the ion
beams are not quite balanced, as illustrated in the (extreme) example
in Figure~\ref{fig:exampleDistribution}. Further, the finite
velocities of the electron holes mean that they are likely closer to
one of the beams in velocity than to the other, and those
velocity-separations evolve as time passes. The situation in the
simulations is thus not quite that of the symmetric cold beams
considered in the proposed hole growth mechanism
\cite{Hutchinson2015}, but the theory nonetheless seems to capture the
main features of the hole growth and beam disruption. Where in the
wake the ion beams are disrupted thus depends on two main factors: the
amount of electron hole growth from the increasing local density, and
the decreasing hole size needed to disrupt the ion beams as their
velocity-separation shrinks.

Further simulations have been carried out to explore the sensitivity
of the ion-beam disruption to changes in mass ratio and electron Debye
length. In general, decreasing the Debye length or the ion to electron
mass ratio leads to earlier disruption of the ion beams. For example,
changing the mass ratio to $\sfrac{m_\mathrm{i}}{m_\mathrm{e}}=115$
leads to the ion beams being disrupted at a time corresponding to
$x\approx 12R_\mathrm{M}$, when they are separated by
$\sim\,$$5.5c_\mathrm{s}$. It is presently not computationally
feasible to do true mass ratio simulations with \espic\ at Debye
lengths much smaller than the $\lambda_\mathrm{De}=0.0025R_\mathrm{M}$
of the simulations presented in this section, so the
$\lambda_\mathrm{De}\sim10^{-5}R_\mathrm{M}$ of the solar wind remains
out of reach. However, the observed trend is for there to be more
electron holes at shorter Debye lengths, appearing and disrupting the
ion beams even earlier in the wake; at
$\lambda_\mathrm{De}=0.00125R_\mathrm{M}$ disruption occurs at
$x\approx 20R_\mathrm{M}$.

\section{Observational Signatures}
\label{sec:observations}

The ARTEMIS mission \cite{Russell2014} has been observing the lunar
wake for several years, and a number of potentially relevant
observations have been presented
\cite{Halekas2011,Tao2012,Halekas2014,Wiehle2011}. Solar wind
variablitity and short Debye lengths in the wake make direct
comparisons with the simulations difficult, but a qualitative
discussion of possible observational signatures of electron holes and
ion-beam disruption is given in this section.

Small electron holes convecting along the dimple in phase space are
one of the two main kinetic phenomena observed in the simulations; the
holes arise because the dimple is unstable \cite{Hutchinson2012}, and
may be present throughout most of the observed wake. Since the holes
remain close to the dimple in phase space their propagation speeds are
of order the electron thermal speed (except very close to the center
of the wake), and the associated pertubations extend $\lesssim$$\,
10\lambda_\mathrm{De}$ in space (parallel to the magnetic field) and
$\lesssim$$\,\sfrac{T_\mathrm{e}}{20e}$ in potential. Such a hole may
thus for instance extend $100\,\mathrm{m}$ along the magnetic field,
be moving with a parallel speed of
$10^6\,\sfrac{\mathrm{m}}{\mathrm{s}}$, and have a potential amplitude
of $0.5\,\mathrm{V}$. That would give an electric field signal of
$\sim$$\,10\,\sfrac{\mathrm{mV}}{\mathrm{m}}$ with duration
$\sim$$\,0.1\,\mathrm{ms}$, which though weaker and faster than those
of electron holes observed in the Earth's plasma sheet
\cite{Andersson2009,Tao2011} may still be detectable with
ARTEMIS. High-frequency electrostatic fluctuations have been observed
in the lunar wake \cite{Tao2012}, and could in principle be poorly
resolved electron holes.

The other main kinetic phenomenon in the simulations is slow-moving
electron holes that grow large enough to disrupt the ion beams; their
typical propagation speeds are of order the sound speed, their
potential pertubations are $\sim$$\,\sfrac{T_\mathrm{e}}{e}$, and
there is no clear upper limit on their sizes. Such holes may be
present beyond a certain distance behind the moon, which depends on
the (variable) solar wind parameters but is expected to fall within
the range covered by ARTEMIS. Since the ion-disrupting electron holes
are slower and perhaps more extended than the small fast holes, their
temporal signatures should be of significantly longer durations. That
could make them easier to detect and characterise, but may also make
them more difficult to disentangle from other signal variations. One
of the high-resolution observations reported by Tao et
al. \cite{Tao2012} revealed broad-band electrostatic fluctuations at
around the ion plasma frequency, for which electron holes were
proposed as one possible cause (cf.
Ref.~\onlinecite{Kojima1997}). Though no well-defined solitary
structures were identified, the fluctuations could be a sign of
ion-disrupting holes, or perhaps of small holes spawned close enough
to the center of the wake to be slow.

Halekas et al. \cite{Halekas2011} presented wake-scale data for the
same traversal studied by Tao et al. Ubiquitous ion-disrupting holes
could in principle affect the ion distribution even at that scale, but
the extent to which such holes could have developed only
$\sim$$\,3.5R_\mathrm{M}$ downstream of the moon is not known from the
simulations. The observed ion spectrogram is dominated by the solar
wind flow, but does show signs of counterstreaming ions. A
higher-resolution segment (Figure~4 of Ref.~\onlinecite{Halekas2011})
could be interpreted as showing the gap between the ion beams closing
(the beams being disrupted) in some regions, but solar wind
variability makes direct interpretation difficult \cite{Wiehle2011}.
That the regions with signs of ion-beam disruption lie close to the
center of the wake is suggestive, however, since that is where the
seeds of the ion-disrupting holes originate in the simulations. It is
also where the high-resolution data showed possible signs of
slow-moving electron holes \cite{Tao2012}, and appears to be the
source of an outward-propagating electron beam \cite{Halekas2011};
electron holes generally increase the parallel electron temperature,
and deep ones can appear as beams.

There is now an abundance of ARTEMIS data available, as illustrated by
recent work \cite{Zhang2014} creating a 3D map of the typical
large-scale structure of the lunar wake. Possible signs of electron
holes have been discussed in this section, but specific searches and
detailed analysis of high-resolution data are needed to confirm the
presence of such holes in the lunar wake.

\section{Conclusions}
\label{sec:conclusions}

The present work highlights the importance of using a realistic ion to
electron mass ratio in 1D simulations of magnetized supersonic wakes
behind non-magnetic objects. Doing so has revealed a novel phenomenon
where the velocity extents of electron holes grow from the small scale
associated with electron--electron instability to one where the holes
can significantly perturb the ions. As a result, the dimple in the
electron distribution remains important despite its narrow width at
realistic mass ratios, acting as a seed of electron holes which
eventually disrupt the counterstreaming ion beams in the wake much
earlier than in the Boltzmann-electron simulations.

Prior kinetic simulations \cite{Birch2001b,Birch2002} observed strong
kinetic electron effects at a highly artificial mass ratio
($\sfrac{m_\mathrm{i}}{m_\mathrm{e}}=20$), but could not establish the
importance of those effects at physical mass ratios because of the
strong mass-ratio dependence of the dimple \cite{Hutchinson2012}. The
present results demonstrate unequivocally the importance of kinetic
electron effects in electrostatic 1D simulations, thus calling into
question the validity of the hybrid simulations frequently used in
attempts to explain experimental observations. It is conceivable that
other physical effects suppress the kinetic electron effects: for
example, magnetic perturbations have been shown to be more important
for solar wind flow which is far from perpendicular to the magnetic
field \cite{Travnicek2005}, and so could in principle interfere with
the parallel electron and ion dynamics. Unless such a suppressing
mechanism can be identified, however, accounting for kinetic electron
effects on ion stability appears to be essential.

Because of the small size of the ion thermal gyroradius compared with
the radius of the moon, the 1D treatment is likely resonably accurate
for the early large-scale structure of the wake. However, the Debye
length is significantly smaller than even the electron thermal
gyroradius, so the accuracy of the 1D non-linear instability evolution
is less certain. In principle the electrostatic modes present in 1D
could couple to electromagnetic modes, possibly at oblique angles,
giving the instabilities a multi-dimensional nature and introducing
pitch-angle scattering, which could affect the electron
dimple. Further, the growing electron holes may not be stable for the
true ordering of length scales, possibly interfering with the ability
of electron holes to reach the size needed to disrupt the ion
beams. Kinetic simulations of higher dimensionality are thus needed,
though these will be computationally challenging given the need to use
a realistic mass ratio and a short Debye length.

Though the lunar wake has been used as the main example of possible
applicability of the present simulations, they solve the general
problem of supersonic flow past a non-magnetic object. As such they
apply equally well to probes in magnetized plasmas and other large
bodies in the solar wind, provided the main assumptions hold. The
importance of electron kinetic effects at supersonic flow also raises
the question of the extent to which similar phenomena can occur at
slower magnetized plasma flow, but to address that question will
require multi-dimensional simulations since the finite extent of the
object in the flow direction and its boundary conditions then matter.


%
%

%

\begin{acknowledgments}
  Useful discussions with David Malaspina regarding electron holes in
  ARTEMIS data are gratefully acknowledged, as is the anonymous
  referee for leading us to add Section~\ref{sec:observations}.
  C.~B.~Haakonsen and C.~Zhou were supported by NSF/DOE Grant
  No.~DE-SC0010491. Computer simulations using \espic\ were carried
  out on the MIT PSFC parallel AMD Opteron/Infiniband cluster Loki.
\end{acknowledgments}

\bibliography{plasma}

\end{document}